# Laser induced breakdown spectroscopy for multielement analysis of powder materials utilized in additive technologies

Lednev V.N.[1,2]*, Sdvizhenskii P.A.[2], Grishin M.Ya.[1,3], Davidov M.A.[1], Stavertiy A.Ya.[4], Tretyakov R.S.[4], Taksanc M.V.[4], Pershin S.M.[1]

[1]*Prokhorov General Physics Institute, Russian Academy of Science, Moscow, Russia*

[2]*National University of Science and Technology MISiS, Moscow, Russia*

[3]*Moscow Institute of Physics and Technology (State University), Dolgoprudny, Moscow Region, Russia*

[4]*Bauman Moscow State Technical University, Moscow, Russia*

*corresponding author's email: vasilylednev@gmail.com

## Abstract

*Powder materials utilized in additive technologies were quantitatively analyzed by laser induced breakdown spectroscopy for the first time. Laser induced breakdown spectroscopy mapping of loose metal powder attached to the double-sided adhesive tape provided high reproducibility of measurements even for powder mixtures with large difference of particles densities (tungsten carbide particles in nickel alloy powder). Laser induced breakdown spectroscopy analytical capabilities for tungsten and carbon analysis were estimated by calibration curve construction and accuracy estimation by leave-one-out cross-validation procedure. Laser induced breakdown spectroscopy and X-ray fluorescence spectroscopy techniques comparison revealed better results for laser induced breakdown spectroscopy analysis. Improved accuracy of analysis and capability to quantify light elements (carbon etc.) demonstrated large perspectives of laser induced breakdown spectroscopy as a technique for express on-site multelement analysis of powder materials utilized in additive technologies.*

## INTRODUCTION

Additive technology bloom challenges analytical chemistry for developing new instrumentation for express on-site characterization of raw powder material and synthesized samples. Production of high quality products by additive technologies implies a good knowledge of powder materials characteristics. Numerous techniques have been already utilized for extrinsic and intrinsic powder properties [1,2]. Concerning chemical analysis, additive powder materials can be directly analyzed by X-ray spectrometry techniques (X-ray diffraction, energy dispersive X-ray spectroscopy (EDX), X-ray photoelectron spectroscopy (XPS), X-ray fluorescence spectrometry (XRF)) or samples can be diluted in acids with subsequent analysis by conventional instrumental techniques (inductively coupled plasma atomic emission spectrometry (ICP-AES), ICP-MS, etc.) [3]. The acid dilution procedure is time consuming thus X-ray spectrometry techniques became conventional and well established procedures characterizing powder materials in additive technologies [2]. However, conventional X-ray techniques are expansive and requires laboratory conditions so they can't be adapted to production lines for express and on-site elemental analysis of powder materials though such demand is of great importance for additive technologies.

Laser induced breakdown spectroscopy is an express multielemental analytical technique for analysis of almost any sample in any environment [4–6]. LIBS technique utilized powerful laser pulse to induce plasma providing emission spectrum which is further analyzed according to atomic/ionic lines. LIBS is a laser based analytical technique quantifying elemental composition with high lateral resolution (typically, 10-50 µm) including depth profile analysis without any sample preparation [7–10]. The remarkable LIBS feature is a capability to perform simultaneous chemical mapping of almost all elements including light elements as carbon, boron or silicon.

LIBS was already demonstrated as a unique tool for online multielement quantitative analysis in industry, including molten steel [11,12] and slags analysis [13], or real time controlling of ore quality at conveyer belt [14]. LIBS technique was successfully adopted for express multielemental analysis of impurities in pharmaceutical powder materials [15–17]. Recently, it was demonstrated that laser induced breakdown spectrometry can be utilized for express multielement analysis of synthesized laser clad coating [18]. However, metal powder analysis is significantly more challenging due to the complexity of numerous powder characteristics influence on results of quantitative analysis.

In this study the results on LIBS quantitative analysis for major elements in metallic powder materials are presented. To the best of our knowledge, LIBS analysis of powder materials utilized in additive technology field have not been published in literature so far. Feasibility of multielemental quantitative LIBS analysis for powder materials within a few minutes has been demonstrated. Generally, in additive technologies powders are mixed during synthesis but pre-mixing is also widely used [19,20]. Nickel alloy and tungsten carbide powders were chosen for analysis in order to demonstrate the perspectives of LIBS due to following reasons. First, nickel alloy reinforced with tungsten carbide particles resulted in composite material with superior wear resistance properties. Such material is of high importance in mining industry and machinery due to possibility of express broken parts reconstruction by additive technologies. Second, carbon is a key element for tungsten carbide coating quality due to the possible chemical interaction with binding matrix elements and formation of complex carbides. Analytical control for carbon is highly required but conventional X-ray techniques (X-ray fluorescence spectroscopy, Electron energy dispersive X-ray spectroscopy) utilized in additive technologies failed to quantitatively analyze light elements (carbon, boron, etc.). Third, tungsten carbide particles are 3-fold heavier compared to 1540-nickel alloy resulting in significant fractionation and poor analytical results by X-ray fluorescence spectroscopy for major components (tungsten, cobalt etc.).

**EXPERIMENTAL**

Quantitative elemental analysis of tungsten and carbon in additive powders utilized for production of wear resistant coatings (nickel alloy reinforced with tungsten carbide particles) [21] was the primary goal of the present study. Powders were purchased from Hoganas Inc. and its chemical compositions are presented in Table 1. In order to estimate LIBS analytical capabilities a series of reference samples were prepared by mixing of known masses (Acculab ALC-210d4) of nickel alloy (1540) and tungsten carbides (WC) powders. Typical WC mass fraction for such wear resistant coating is in 20-50 % range [19,20]. Five samples (20, 25, 30, 35, and 45 wt. % of tungsten carbide in 1540 alloy) were prepared. All reference samples were analyzed by compact X-ray fluorescence spectrometer (Innov-X Delta, Olympus). Powder samples were put inside polyethylene cuvette with polyester window and extensively mixing by shaking before measurements. Examples of scanning electron microscopy images (JEOL, JSM-646-LV) for WC and 1540 powders are presented in Fig. 1. Tungsten carbide particles had mostly sphere shape of 70-80 μm diameters while few particles composed by 10 μm grains aggregates while 1540 alloy particles were slightly bigger in size (70-120 μm).

Laser induced breakdown spectroscopy measurements were performed with the custom-made setup [22,23] based on a pulsed solid state Nd:YAG laser (1064 nm, 10 ns, 0.5 mJ/pulse, 5 Hz, $M^2$=2). Laser beam was focused normally through microobjective (x4, focal length 20 mm). The choice of such tight focusing was made in order to obtain small laser spot (diameter 50 μm) which should be smaller compared to typical diameter of tungsten carbide (WC) and nickel alloy (1540) particles. Laser plasma was generated in air, plasma emission was collected according to side-view scheme by quartz lens (F=60 mm) and transferred to the entrance slit of spectrometer (Shamrock 303i, Andor) equipped with gated detector (iStar, Andor). Sample holder was installed on two-dimensional motorized stage which can be moved with 0.1 μm step and 0.2 μm position precision. Control and synchronization of motorized stage, laser and spectrometer was carried out by custom-made software developed in LabVIEW environment.

**RESULTS AND DISCUSSION**

The optimal choice of LIBS spectral lines depends on a number of factors including detector sensitivity, line intensity and spectral interference. The choice of analytical line for carbon is challenging in LIBS due to strong spectral interference with tungsten, nickel, chromium and iron lines. The strongest carbon line C I 247.86 is spectrally interfered with iron Fe I 247.86, nickel Ni I 247.70 and strong tungsten W II 248.14 lines. Carbon line C I 833.51 is spectrally interfered with iron line Fe I 832.70 and tungsten line W I 832.23 [24]. Consequently, a carbon line C I 193.09 was chosen as analytical line. Laser induced plasma spectra for loose powder of tungsten carbide and 1540-nickel alloy are presented in figure 2. Chosen spectral window 189-210 nm contained atomic/ionic lines for all major components. The following lines for nickel (Ni I 205.99) chromium (Cr I 199.99), iron (Fe I 193.45) and tungsten (W II 207.91) were chosen due to absence of spectrally interference and sufficiently high intensity. Plasma emission duration was less than 2 μs thus a 1 μs gate and a short delay of 0.2 μs was used to suppress continuum emission during first moments of plasma expansion. Firstly, direct single shot ablation of loose powder in a cavity was utilized. However this resulted in ejection of particles and formation of cone crater (diameter 3 mm, depth 2 mm) thus decreasing reproducibility of sampling. Moreover, significant difference of WC and 1540-alloy densities (12.5 vs 4.7 g/cm$^3$ according to supplier specification) leads to preferential enrichment of heavy WC particles at laser crater bottom. In order to diminish such influence alternative sampling procedures were suggested.

Express and on-site LIBS analysis was a primary goal in the current study so sampling method should be simple, effective and robust for metal powder analysis: gluing in epoxy, loose powder on adhesive tape and pressed powder in soft metal foil. Tungsten carbide and 1540 alloy particles density differed 3-fold so sampling procedure should be capable to provide uniform distribution of different particles. First procedure was based on powders mixture homogenization in viscous liquid which became stable with time, i.e. epoxy glue. The second method of sampling utilized double sided adhesive tape attached to microscope slide and loose powder spreading on top of the tape. Excess powder was discarding by tapping the slide, leaving the uniform thin layer of powder on the tape. Double sided polyethylene film and foamed polyethylene adhesive tapes were used for powder caption. Third sampling procedure utilized powder pressing into soft metal foil (i.e. chemically clean copper, 99.9 % wt.) thus avoiding introduction of carbon. All sampling procedures were compared with focusing on carbon pollution prevention. For better identification of carbon pollution another additive material powder (inconel 625) was used as a test material due to low concentration of carbon (Table 1). Examples of LIBS spectra acquired for all three sampling procedures and corresponding samples surface photos are presented in Fig. 3. A series of 900 LIBS spectra was acquired and summed for every laser shot achieving a fresh surface (30x30 spots). "Epoxy gluing" procedure resulted in 12-fold increasing of carbon line C I 193.09 compared to "adhesive tape" or "copper foil" sampling. Pressing in copper foil introduced small amount of carbon but further experiments with WC/1540 alloy mixtures demonstrated that uniform distribution of particles is difficult if possible to achieve with such sampling procedure. Additionally, particles surface was slightly damaged during pressing (Fig. 3 a) so grains can be polluted by elements from press punch. "Adhesive tape" procedure was chosen as sampling procedure due to low carbon pollution, capability to obtain uniform particle distribution and simplicity of sampling. Foamed adhesive tape sampling provided lower intensity of carbon line C I 193.09 because of more dense particles distribution resulting in low possibility to ablate adhesive tape between particles (Fig. 3).

Poor reproducibility of LIBS spectra was observed during first experiments because of significant difference in densities for WC and 1540-alloy particles (12.5 vs 4.7 g/cm$^3$). In order to improve homogeneity of WC and 1540 particles distribution a simple and effective sampling procedure was carried out. Reference samples were placed in 100 ml polyethylene box closed with a cap and then it was shaken with different amplitudes (2-30 mm) at 5 Hz frequency rate during 30 seconds. Double sided foamed adhesive tape was attached to flat surface of aluminum cube to form "a sampling head". This head (mass 50 g) was put at the loose powder surface inside mixing box thus upper layer of particles was glued to the adhesive tape. Then adhesive layer with attached particles was directed downwards in order to discard excess particles by tapping the other side of sampling head. Sampling head was placed in sampling holder insuring constant lens-to-sample distance and which was attached to two-dimensional motorized stage for obtaining LIBS measurements in scanning maps mode.

"Strong" (amplitude >20 mm) and "gentle" (amplitude <3 mm) shaking hominization procedures with following "adhesive tape sampling" were compared quantitatively. To do so 20 % WC reference sample was mapped (6x6 mm area) for tungsten line W II 207.91 by LIBS.

Particles dimensions were in range 70-120 μm and laser crater was of 50 μm diameter so areas were mapped with 200 μm step in order to eliminate possible influence of previous craters formation. Five parallel measurements of W II 207.91 maps were obtained for "strong" and "gentle" mixing and homogenizing reproducibility was estimated as relative standard deviation for "tungsten pixels" which was determined as a sum of spots with W II 207.91 line integral greater than 500 counts. Examples of first maps for "strong" and "gentle" mixing and corresponding tungsten particles number for replicate measurements were presented in Fig. 4. "Strong" mixing provided good reproducibility of sampling (RSD = 6.7%) which was 8-fold better compared to "gentle" mixing where significant fractionation was taking the place.

The larger sampled area the better LIBS analysis representative in case of non-homogeneous materials analysis. In order to optimize time required for single LIBS measurement and estimate minimum sampling area for representative analysis a loose powder mixture on adhesive tape was mapped for 8x8 mm area with 0.2 mm spatial resolution (1600 sampling spots). Sample with lowest tungsten carbide concentration (WC 20 % wt.) was used since this sample is most challenging for representative analysis. Results of Ni I 205.99 and W II 207.91 lines integrals mapping are presented in Fig. 5. It is clearly seen that tungsten carbide particles (red) are randomly dispersed at the surface. Note, that almost 6% of spots were marked with white color because of too low intensity for Ni I 205.99 or W II 207.91 lines that was explained by adhesive tape ablation (area between particles) or at the particle edge surface ablation. In order to estimate minimal sampling area for representative analysis Fig. 5 (a) was analyzed in the following way. Tungsten to nickel spots ratios for squared areas of different sizes were calculated. For example, for 10x10 spots area the tungsten and nickel pixels were counted excluding pixels with low intensities (less 100 a.u. in Fig. 5a) and corresponding W/Ni ratio was calculated. The same W/Ni ratios was estimated for sampling areas of 2x2 (4 pixels), 3x3 (9 pixels) and up to 40x40 (1600 pixels) sampling spots and this ratio was plotted as a function of sampling spots number (Fig. 5b). W/C ratio fluctuated significantly during first 500 sampling spots but then its reproducibility (relative standard deviation) improved and did not exceed 5 %.

Calibration curves were constructed by replicate measurements of five different sampling areas with 30x30 spots (900 measurements each). Tungsten W II 207.91 and carbon C I 193.09 line integrals with background correction were normalized on Ni I 205.99 line and then plotted as a function of tungsten and carbon mass fraction (Fig. 6). Calibration curve for tungsten was fitted with linear function while carbon calibration curve clearly demonstrated saturation and was fitted by quadratic function. In both cases quality of fits were rather high with R-squared ($R^2$) greater than 0.990. In order to evaluate the accuracy of detection a leave-one-out cross-validation procedure with the root mean square of cross-validation (RMSECV) as a metric was utilized [8]. The RMSECV was defined with the following equation:

$$RMSECV = \sqrt{\frac{1}{n}\sum_{i=1}^{n}(c_i - \hat{c}_i)^2}$$

where *n* is the number of samples, $c_i$ is the predicted concentration of sample and $\hat{c}_i$ is the reference concentration of the sample. The basic idea of RMSECV is to leave one sample from a set, redraw the calibration curve and predict the concentration of this point. Then take the full set of samples, leave another point from calibration curve, redraw it and calculate the concentration of this point and then repeat such procedure for every point in a set. The estimated RMSECV values for tungsten and carbon (Fig. 6) were fairly good (1.30 and 0.122 % wt. respectively).

Furthermore, we've compared LIBS results with data acquired with conventional technique utilized for analysis of powders in additive technology: X-ray fluorescence spectroscopy. Example of XRF results and comparison with LIBS are presented in Fig. 7. A conventional sampling procedure with filling the polyethylene cuvette equipped polyester window by reference samples was utilized. Five XRF parallel measurements were made to extract mean and reproducibility (standard deviation). Unfortunately, XRF results fluctuated a lot (Fig. 7 a) due to 3-fold difference of WC and 1540-alloy particle density resulting in non-homogeneous distribution of tungsten carbide particles. Comparison of XRF and LIBS results (Fig. 7 b) demonstrated that last technique provided better accuracy of analysis. Additionally, LIBS technique was capable to quantitatively analyze light elements (carbon. boron, etc.) which are of great importance for additive materials but cannot be analyzed by XRF.

CONCLUSIONS

A feasibility of laser induced breakdown spectroscopy (LIBS) for quantitative analysis of metal powders utilized in additive technologies was demonstrated for the first time. A simple and effective sampling procedure of loose metal powder on double sided adhesive tape was utilized. LIBS mapping revealed that uniform grains distribution can be achieved for adhesive tape sampling despite 3-fold difference of WC and 1540-alloy particles densities. Sampling area dimensions was optimized to fulfill requirements for representative LIBS analysis while minimizing time needed for LIBS measurements. LIBS analytical capabilities for tungsten and carbon analysis were estimated by calibration curve construction with focusing on linearity and root mean square of cross-validation (RMSECV) metrics. LIBS results for tungsten analysis was better than for XRF measurements due to better reproducibility of sampling procedure. LIBS technique was also capable to quantitatively analyze carbon in metal powders. Current study revealed bright perspectives of LIBS for express on-site analysis of powder materials utilized in additive technologies.

FUNDING

The authors gratefully acknowledge the financial support of the Russian Science Foundation (agreement № 16-19-10656).

**Table 1**. 1540-nickel alloy, tungsten carbide and Inconel 625 powders chemical composition and properties.

| Powder | Fe | Ni | C | Mo | Cr | Mn | Si | B | Nb | W |
|---|---|---|---|---|---|---|---|---|---|---|
| **1540** | 2.36 | base | 0.27 | - | 7.55 | - | 3.51 | 1.64 | - | - |
| **WC** | 0.2 | | 4 | | | | | | | base |
| **Inc 625** | 0.67 | base | 0.10 | 8.9 | 21.3 | 0.39 | 0.42 | - | 3.58 | - |

**Figure 1** Scanning electron microscopy images of tungsten carbide (a) and 1540 nickel alloy (b) particles.

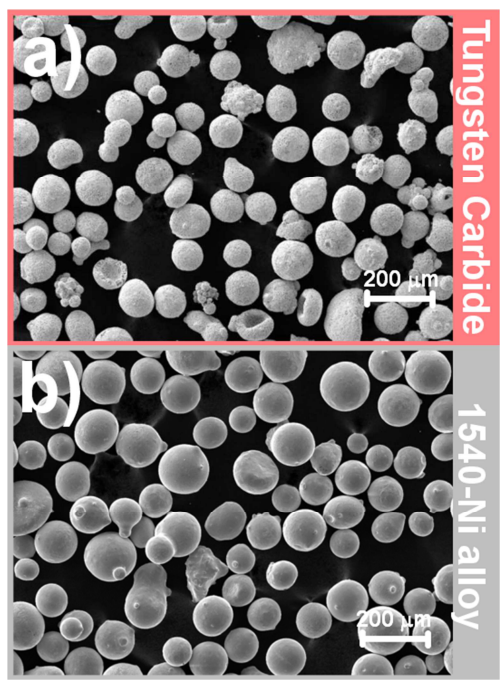

**Figure 2.** Laser induced breakdown spectra for tungsten carbide (red) and 1540-alloy (black) powders. Spectra were acquired with 1 μs gate and 0.2 μs delay. Tungsten carbide spectrum was vertically shifted by 5*10$^3$ counts for better view.

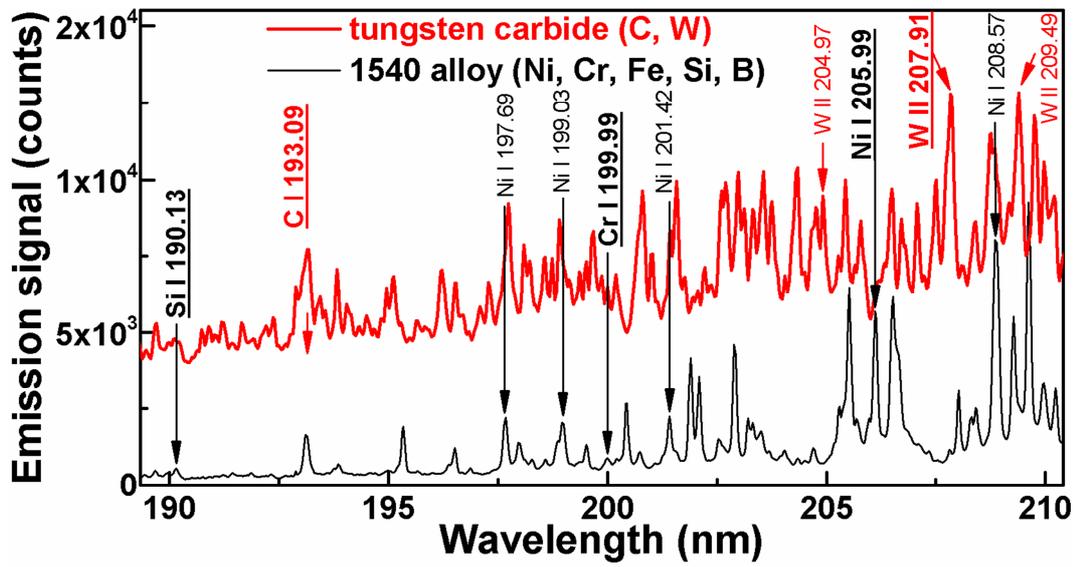

**Figure 3.** Sampling area images (a) and corresponding laser induced plasma spectra (b) for sampling of inconel 625 powder in epoxy (black), glued to polyethylene (cyan) and foamed (violet) double-sided adhesive tape and pressed in copper foil (red). Spectra were multiplied and then vertically shifted (copper foil – by $5*10^6$ counts, adhesive tape – by $1*10^7$ counts, epoxy - by $1.5*10^7$ counts) for better comparison.

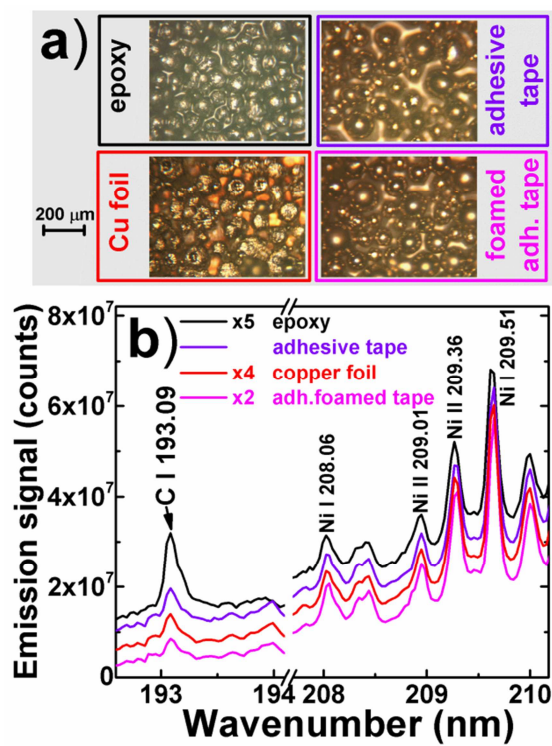

RSD - relative standard deviation.

**Figure 4.** Emission signal maps of W II 207.91 line for "strong" (a) and "gentle" (b) homogenized procedures followed by "adhesive tape" sampling. Five parallel measurements (five sampling areas) were mapped for "strong" and "gentle" mixing procedure and (a) and (b) are examples of the first maps. Homogenization reproducibility (c) was estimated as a relative standard deviation (RSD) of tungsten spots number (tungsten spot was defined as spot with W II 207.91 integral greater than 500 counts).

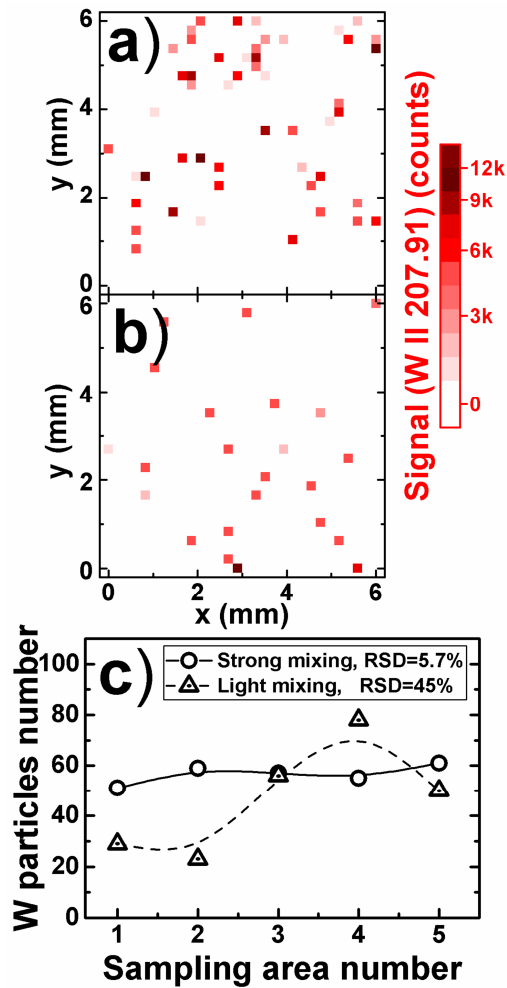

**Figure 5.** Optimizing sampling area dimensions for "adhesive tape" procedure. Nickel Ni I 205.99 and tungsten W II 207.91 lines integrals (a) with background correction were mapped for 8x8 mm area covered with mixture of tungsten carbide and 1540-Ni alloy on adhesive tape (laser spot 50 μm, distance between sampling spot 0.2 mm, 1600 measurement spots). Note that almost 8% spots are marked with white color because of too low signals for Ni I 205.99 or W II 207.91 lines due to low ablation at particle side plane or adhesive tape. Tungsten-to-nickel spots ratio (b) averaged by increasing area dimensions or sampling spot number. Dashed lines in (a) are examples of squared areas (10x10 square = 100 spots, … 40x40 square = 1600 spots) utilized for counting W/Ni spots ratio.

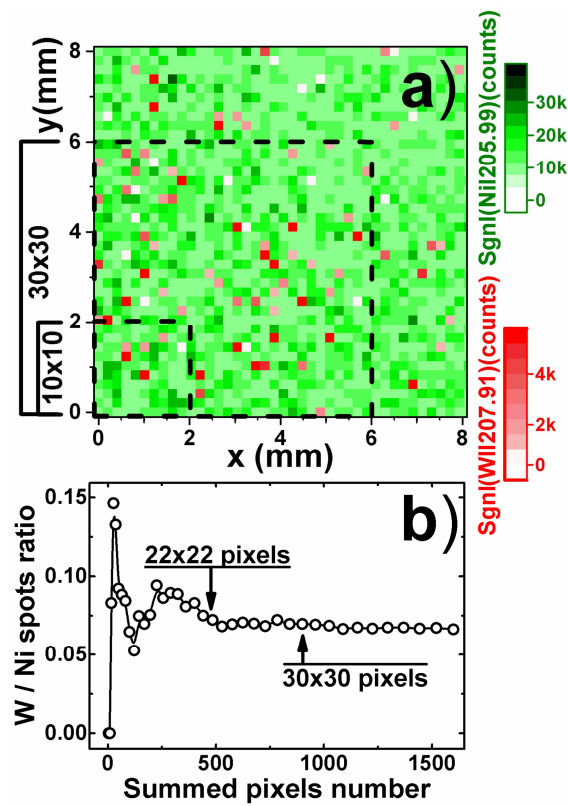

$R^2$ – coefficient of determination (R-squared); RMSECV - root mean square of cross-validation; wt. % – mass fraction expressed in weight percent.

**Figure 6.** Laser induced breakdown spectroscopy calibration curves for tungsten (a) and carbon (b) for reference samples. Tungsten and carbon concentrations are presented as mass fraction (weight %). Calibration curves were fitted with linear (a) and quadratic (b) functions and were compared in terms of coefficient of determination ($R^2$) and root mean square of cross-validation (RMSECV) metric.

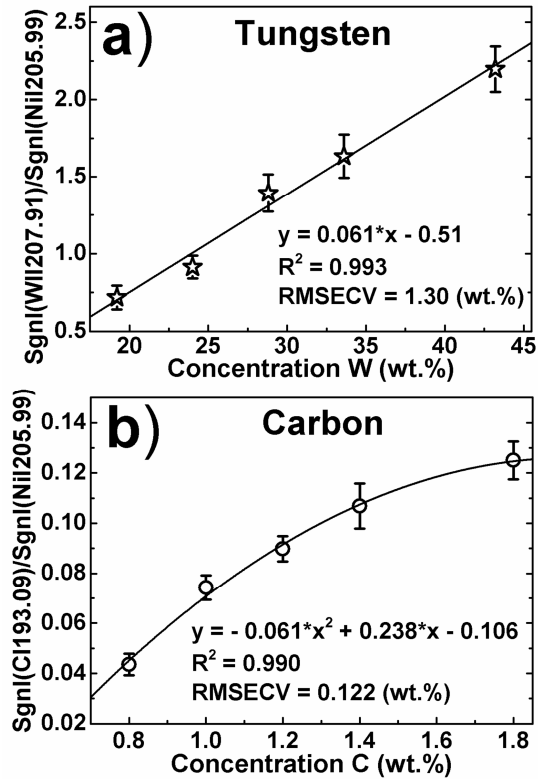

XRF – X-ray fluorescence spectroscopy; LIBS – laser induced breakdown spectroscopy; wt. % – mass fraction expressed in weight percent.

**Figure 7.** X-ray fluorescence (XRF) spectroscopy analysis vs calculated tungsten mass fraction according to reference samples (a). Laser induced breakdown spectroscopy (LIBS) and X-ray fluorescence spectroscopy (XRF) results comparison (b) for quantitative analysis of tungsten.

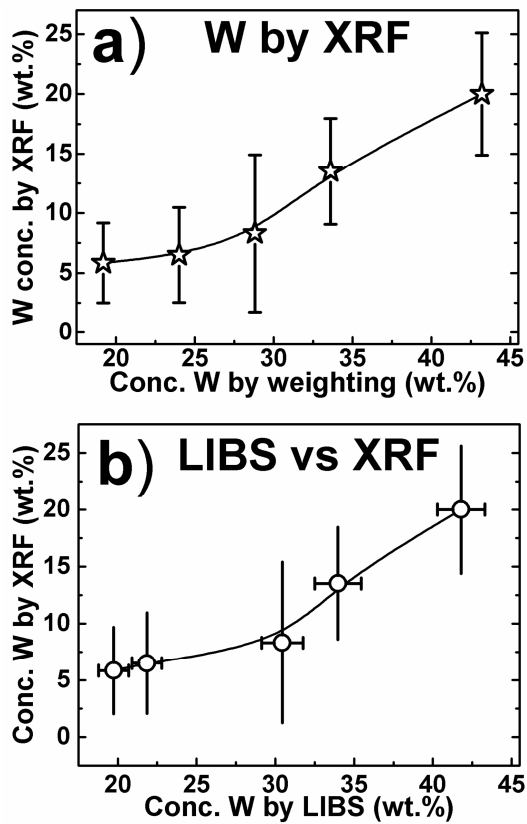